\documentclass[12pt]{article}
\usepackage{amsmath}
\usepackage{amssymb}
\usepackage{amscd}

\begin{document}
\begin{center}
\large\bf
Cryptanalysis and improvements on some\\ graph-based authentication schemes\\\
\end {center}

\begin{center}
\small
  Herish O. Abdullah\\
 Salahaddin University, College of Science\\ Mathematics Department, Kurdistan, Iraq\\
 email:herish\_omer69@yahoo.com\\ 
 Mohammad Eftekhari\\
LAMFA, CNRS UMR 7352\\
 Universit\'e  de Picardie-Jules Verne\\33 rue Saint-Leu 80039 Amiens France\\
email:mohamed.eftekhari@u-picardie.fr\\

\end{center}

\begin{center}
{\large\bf
Abstract}
\end{center}

{\it In 2010, Grigoriev and Shpilrain, introduced  some graph-based authentication schemes. We present a cryptanalysis of some of these protocols, and introduce some new schemes to fix the problems.\\

}\\
\vfill

{\bf\large 1. Introduction}\\
 
We refer the reader to [12,19] for details of the general theory of public-key authentications and the terminology of Feige-Fiat-Shamir-like authentication schemes. The concept of Feige-Fiat-Shamir authentication scheme was introduced in 1987[5]. It is a zero knowledge proof, which is a procedure for a prover (Alice) to convince a verifier (Bob) that a fact is true without revealing anything other than the veracity of the fact to be proven.\\

In [3,4], graph theory was used for the first time to construct such authentication schemes, in contrast to the existing ones which were based on number
theoretical problems. However in [3,4], it is not clear how the scheme works and why it is secure.\\
In [6], Grigoriev and Shpilrain proposed several general Feige-Fiat-Shamir-like authentication schemes. They employed some NP-hard problems in graph theory such as graph homomorphism problem(GHP), subgraph isomorphism problem(SGIP) and graph coloring problem(GCP) to construct some platforms.\\
After the knapsack cryptosystem was broken, combinatorial-based cryptography fell into disfavor (see [10] for an interesting discussion); the above proposals
are interesting in the sense they try to reintroduce combinatorics in cryptography. In this paper we cryptanalyse some of the proposals of Grigoriev and 
Shpilrain, pointing out some security problems and we propose some new schemes fixing these problems.\\

The paper is organized as follows:\\
In 2.1  we recall the authentication protocol based on graph homomorphism problem. In 2.2 we present a cryptanalysis of this scheme. Using this attack the adversary will be able to impersonate the prover and fool the verifier.
Then in 2.3, we propose a new protocol in the same spirit,  based on subgraph isomorphism problem which resists the attack of 2.2.
 In 3.1 we recall another protocol of [6], based on the graph coloring problem. In 3.2 we point out  a security weakness in this  platform resulting from an indirect use of graph isomorphism problem.  Moreover, in 3.3 we propose a new protocol in the same spirit which fixes this weakness.\\
\vfill

{\bf\large 2. Cryptanalysis against graph homomorphism based protocol}\\ 

In this section, we review graph homomorphism based protocol proposed by Grigoriev and Shpilrain and present a cryptanalysis against the platform. Also, we propose a new protocol to fix the problem.\\

{\bf\large 2.1 GH-based protocol}\\

Given two graphs $\ G$ and$\ H$, a homomorphism from $\ G$ to$\ H$ is a mapping $\varphi:\ G\to \ H$ that satisfies the following: $\ uv \in\ E(G)\Rightarrow\varphi (u)\varphi (v)\in\ E(H)$. The Graph Homomorphism Problem(GHP) asks whether or not there is a homomorphism $\varphi$ from $\ G$ onto $\ H$. In 1990, Hell and Nesetril[7] showed that the GHP is NP-complete unless $\ H$ 
has a loop or is bipartite.\\
 
We recall the graph homomorphism-based authentication protocol proposed in [6]. \\
Alice's public key consists of two graphs, $\Gamma_1$ and $\Gamma_2$, and her private key is a surjective homomorphism $\alpha:\Gamma_1\to \Gamma_2$. To begin authentication:\\
1. In the commitment step, Alice picks a graph $\Gamma$ together with a surjective homomorphism $\beta:\Gamma\to \Gamma_1$, and sends $\Gamma$ to Bob, while keeping $\beta$ secret.\\
2. Bob sends Alice a random bit $\ b \in\{0,1\}$, the challenge.\\
3. If$\  b=0$, Alice sends Bob the homomorphism $\beta$, and if $\ b=1$, then she sends the composition $\alpha o\beta$.\\
4. If$\  b=0$, Bob verifies whether $\beta(\Gamma)= \Gamma_1$ and whether $\beta$ is a homomorphism; and if $\ b=1$, then he verifies whether $\alpha\beta(\Gamma)=\Gamma_2$ and whether $\alpha o\beta$ is a homomorphism or not.\\

The security of this protocol is based on the difficulty of finding a homomorphism $\alpha$ from $\Gamma_1$ onto $\Gamma_2$. \\

As mentioned by the authors, an eavesdropper doesn't need to discover the secret keys of Alice to attack this protocol. In fact, if he can find a graph, say $\Gamma'$ that maps homomorphically onto $\Gamma_1$ and onto $\Gamma_2$, say $\alpha':\Gamma'\to\Gamma_1$ and $\beta':\Gamma'\to\Gamma_2$, then he can interfere in the commitment step and respond to either challenges of Bob.\\

In the next subsection, we use this idea to attack the above protocol.\\

{\bf\large 2.2 Cryptanalysis of the GH-based protocol}\\

In this subsection, we present a forgery attack against the GH-based protocol by using the fact that the tensor product is the category-theoretic product in the category of graphs and graph homomorphisms. The tensor product $\Gamma_1\otimes \Gamma_2$ of graphs $\Gamma_1$ and $\Gamma_2$ is a graph with vertex set $\ V(\Gamma_1)\times V(\Gamma_2)$ where two vertices $\ (u_1,v_1)$ and $\ (u_2,v_2)$ are adjacent when $\ u_1 u_2\in E(\Gamma_1)$ and $\ v_1 v_2\in E(\Gamma_2)$[8, p.163]. \\

Now, let $\Gamma'=\Gamma_1\otimes\Gamma_2$, and define $\alpha':\Gamma'\to\Gamma_1$ and $\beta':\Gamma'\to\Gamma_2$ by $\alpha'(u,v)=u$ and $\beta'(u,v)=v$, for each vertex $\ (u,v)\in\ V(\Gamma_1\otimes\Gamma_2$).\\

One can easily check that  $\alpha'$ and $\beta'$   are homomorphisms.\\

From this we conclude that, an eavesdropper(Charlie) can successfully interfere at the commitment step and respond to either challenges of Bob as follows:\\
1. Charlie computes $\Gamma'=\Gamma_1\otimes\Gamma_2$,  $\alpha'(u,v)=u$ and $\beta'(u,v)=v$, for each vertex $\ (u,v)\in\Gamma'$ and sends $\Gamma'$ to Bob.\\
2. Bob sends Charlie a random bit $\ b \in\{0,1\}$.\\
3. If$\  b=0$, then Charlie sends Bob the homomorphism $\beta'$, and if $\ b=1$, then he sends $\alpha'$.\\
4. If$\  b=0$, then Bob verifies whether $\alpha'(\Gamma')= \Gamma_1$ and whether $\alpha'$ is homomorphism; and if $\ b=1$, then he verifies $\beta'(\Gamma')=\Gamma_2$ and whether $\beta'$ is a homomorphism.\\

According to the above procedure, the verifier will be convinced by Charlie that he knows the secret keys, hence an eavesdropper will be able to fool the verifier. We conclude that the GH-based protocol is completely impractical.\\

Next, we propose a new protocol to fix this problem.\\

{\bf\large 2.3 New proposed protocol based on subgraph isomorphism problem}\\

In this subsection, we present a new protocol based on the subgraph isomorphism problem to fix the above problem. Two graphs $\ G_1$ and$\ G_2$ are $\ isomorphic$ if there is a one-to-one and onto mapping $\varphi:\ G_1\to \ G_2$ that preserves adjacency( and non adjacency), that is $\ uv \in\ E(G_1)$ if and only if $\varphi (u)\varphi (v)\in\ E(G_2)$, for any two vertices $\ u$,$\ v\in V(G_1)$.\\

Given two graphs $\ G$ and $\ H$, subgraph isomorphism problem(SGIP) asks whether or not $\ H$ is isomorphic to a subgraph of $\ G$. In 1971, Stephen Cook[16] showed that SGIP is NP-complete. Below we give a description of our proposed protocol based on SGIP.\\

Alice's public key consists of two graphs $\Omega$ and $\Gamma_2$, and her private key is a subgraph $\Gamma_1$ of $\Omega$ together with an isomorphism $\alpha:\Gamma_1\to \Gamma_2$.\\
1. In the commitment step, Alice chooses a graph $\Lambda$ which is isomorphic to a subgraph $\Lambda'$ of $\Omega$ with $\Gamma_1\subset\Lambda'$. She also chooses an embedding $\beta:\Lambda\to\Omega$ ($\beta(\Lambda)=\Lambda'$), and sends the graph $\Lambda$ to Bob, while keeping $\Lambda'$ and $\beta$  secret. Note that there exists a subgraph $\Gamma\subset\Lambda$ with $\beta(\Gamma)=\Gamma_1$.\\
2. Bob sends Alice a random bit $\ b \in\{0,1\}$, the challenge.\\
3. If$\  b=0$, then Alice sends Bob the embedding $\beta$, and if $\ b=1$, then she sends the subgraph $\Gamma\subset \Lambda$ and the composition $\alpha o\beta |_\Gamma$.\\
4.  If$\  b=0$, then Bob verifies whether $\beta$ is an embedding of $\Lambda$ into $\Omega$, and if $\ b=1$, then he verifies whether $\Gamma\subset\Lambda$, $\alpha o\beta |_\Gamma(\Gamma)=\Gamma_2$ and that $\alpha o\beta$ is an isomorphism of $\Gamma$ into $\Gamma_2$.\\ 

{\bf\ Proposition 1.} Suppose that after several runs of the steps of the above protocol, both values of $\ b$ are encountered. Then, successful forgery in the protocol is equivalent to finding a subgraph $\Gamma'_1$ of $\Omega$ together with an isomorphism $\pi:\Gamma'_1\to\Gamma_2$. \\
{\bf\ Proof}. Suppose Charlie wants to impersonate Alice. To that effect, he interferes in the commitment step by sending his own commitment $\Lambda'$ to Bob. Since he should be prepared to respond\\  to the challenge $\ b=0 $, he should know an embedding $\beta':\Lambda'\to\Omega$. On the other hand, since he should be prepared for the challenge $\ b=1$, he should know an isomorphism $\pi:\Gamma'\to\Gamma_2$ with $\Gamma'\subset\Lambda'$. Now, since $\Gamma'$ is isomorphic to a subgraph of $\Omega$ , this implies that he can produce a subgraph $\Gamma'$ of $\Omega$ which is isomorphic to $\Gamma_2$. This completes the proof. $\Box$\\

{\bf\large 3. A Weakness in authentication scheme based on graph coloring}\\ 

In this section, we review graph coloring based protocol proposed in[6] and point out a weakness in the scheme. We also propose a new scheme to fix the problem. \\ 

{\bf\large 3.1 GC-based protocol}\\

Given a connected graph $G$ and a positive integer $ k\le p(G)$, where $ p(G)$ is the order of the graph $G$, a $k-$coloring of $ G$ assigns a color from $\{1,2,...,k\}$ to each vertex of $ G$ so that adjacent vertices recieve distinct colors. In 1972, Karp[15] showed that the graph coloring problem is NP-complete.\\

Grigoriev and Shpilrain proposed a generic protocol whose difficulty is based on "most any" search problem. Then they gave a platform based on the graph coloring problem to illustrate the idea of the protocol. The GC-based scheme is described as follows:\\

Alice's public key consists of a $k$-colorable graph $\Gamma$, and her private key is a $k$-coloring of $\Gamma$, for some (public)$\ k$. To begin authentication,\\
1. In the commitment step, Alice picks a graph $\Gamma_1$ together with an isomorphism $\varphi:\Gamma\to\Gamma_1$, and sends the graph $\Gamma_1$ to Bob, while keeping the isomorphism $\varphi$ secret.\\
2. Bob sends Alice a random bit $\ b \in\{0,1\}$, the challenge.\\
3. If$\  b=0$, then Alice sends Bob the isomorphism $\varphi$, and if $\ b=1$, then she sends a $k$-coloring of $\Gamma_1$.\\
4. If$\  b=0$, then Bob verifies whether $\varphi$ is an isomorphism from $\Gamma$ into $\Gamma_1$; and if $\ b=1$, then he verifies this is indeed a $ k$-coloring of $\Gamma_1$.\\

The authors showed that successful forgery in the above protocol is equivalent to finding a $\ k-$coloring of the graph $\Gamma$. In fact this is not true, and we give a weakness in the above protocol in the next subsection. \\

{\bf\large 3.2 A Weakness in GC-based scheme}\\ 

In this subsection, we point out a weakness in GC-based scheme, resulting from the graph isomorphisms used in the construction of the protocol. Using this weakness, Bob will be able to find out the Alice's secret key.\\

Given two graphs $\ G_1$ and $\ G_2$, the Graph Isomorphism Problem(GIP) asks whether or not there is an isomorphism $\varphi:\ G_1\to\ G_2$.\\

Besides its importance in practice, the GIP is prominent in computational complexity theory as it is one of a very small number of problems belonging to NP neither known to be solvable in polynomial time nor NP-complete. Two other problems which were thought to have the same status have been solved in polynomial time: linear programming problem which was shown to be in P in 1979 by Khachian[9], and the problem of determining the primality of an integer was shown to be in P in 2002 by Agrawal, Kayal, and Saxena[1]. \\

In [2], Babai, Erd\"os and Selkow showed that for almost all graphs X, any graph Y can be easily tested for isomorphism to X by an extremely naive linear time algorithm. \\

Well known efficient algorithms for finding isomorphisms between random graphs are: Nauty algorithm by Brendan Mckay[11] and Nauty's improvements such as Saucy[14] and Bliss[18]. Nauty Algorithm is one of the most efficient and powerful algorithms that solve GIP in polynomial time for random graphs with thousands or more vertices.\\

On the other hand, some researchers tried to find hard graphs for Nauty-like algorithms. In 1997, Miyazaki[13] constructed a family of colored graphs which are hard for Nauty algorithm and  require exponential time. In 2009, Greg D. Tener[17] introduced "nishe-algorithm" that solves Miyazaki graphs in polynomial time.\\ 

Therefore, for many large random graphs we surely can find isomorphisms(if they are isomorphic) by using one of the above algorithms. So if we wish to use GIP in cryptography, we must work with a small family of graphs, and even with this we can not be sure that someone using a combination of the existing algorithms is not able to attack it successfully.\\

Moreover, the most important point for cryptographic security is computational intractability of a problem on a generic set of inputs, i.e the problem should be hard on "most " randomly selected inputs. And this is not the case for GIP.

Henceforth, we conclude that GIP is not suitable for the design of authentication protocols even if it is used in an indirect way.\\

Now, returning to GC-based scheme, we see that Bob can fool Alice as follows:\\
Alice sends Bob a graph $\Gamma_1$ isomorphic to $\Gamma$.\\
$\diamondsuit$ Bob sends Alice the challenge $\ b=1$.\\
$\diamondsuit$ Alice sends Bob a $\ k-$coloring of the graph $\Gamma_1$.\\
$\diamondsuit$ Using an efficient algorithm, Bob will compute an isomorphism $\psi:\Gamma_1\to\Gamma$, and then he easily can deduce from $\Psi$, a $ k$-coloring of the graph $\Gamma$, which is the secret key of Alice.\\

A new protocol based on GCP and SGIP will be given in the next subsection. This protocol fixes the above problem.\\

{\bf\large 3.3 New proposed protocol based on graph coloring problem and subgraph isomorphism problem}\\

As we mentioned in the previous subsection, using graph isomorphism problem is not suitable in the design of authentication schemes. Instead, we will use subgraph isomorphism problem and also we will use an intermediate graph to hide the secret keys as follows.\\

Alice chooses a graph $\Gamma$ which contains a $ k$-colorable subgraph $\Gamma_1$ of order $\ n$. $\Gamma$, $\ n$ and $ k$ are public, while the graph $\Gamma_1$ with a $ k$-coloring of $\Gamma_1$ are secret. To begin the authentication:\\
1. In the commitment step, Alice picks an intermediate graph $\Lambda'$ which is a subgraph of $\Gamma$ and contains the graph $\Gamma_1$ together with an isomorphism $\alpha:\Lambda\to\Lambda'$ and sends $\Lambda$ to Bob.\\
2. Bob sends Alice a random bit $\ b \in\{0,1\}$.\\
3. If$\  b=0$, then Alice sends Bob an embedding $\beta:\Lambda\to\Gamma$, and if $\ b=1$, then she sends a subgraph $\Gamma_2\subset \Lambda$ of order $n$, together with a $ k$-coloring of $\Gamma_2$.\\
4.  If$\  b=0$, then Bob verifies whether $\beta$ is an embedding of $\Lambda$ into $\Gamma$, and if $\ b=1$, then he verifies whether $\Gamma_2$ is a subgraph of $\Lambda$ and that the  $ k$-coloring is indeed a coloring of $\Gamma_2$.\\ 

The following result may be proved in much the same way as Proposition 1.\\

{\bf\ Proposition 2.} Suppose that after several runs of the steps of the above protocol, both values of $\ b$ are encountered. Then, successful forgery in the protocol is equivalent to finding
 a subgraph $\Gamma_1$ of $\Gamma$ of order$\ n$ together with a $ k$-coloring of $\Gamma_1$. \\

{\bf\large 4 Conclusion}\\

We cryptanalyzed two graph-based authentication protocols. For one of them, we showed it is completely impractical, and proposed a new scheme instead.
  For a second protocol, we pointed out a weakness and proposed a new one, solving the problem.
    A detailed complexity study of the above protocols, precising the kind of graphs to be used, the number of nodes etc, has to be done. We hope doing that in the future.\\

\vfill

{\bf\large Bibliography}\\

[1] Agrawal, Kayal, and Saxena, PRIMES is in P, Annals of Mathematics 160 (2), 2002, p. 781-793.\\

[2] L. Babai, P. Erd\"os, and S.M. Selkow, Random graph isomorphism. SIAM Journal on Computing, 9(3), 1980, p. 628-635.\\

[3] P. Caballero-Gil, C. Hern´andez-Goya, Strong Solutions to the Identification Problem, 7th Annual International Conference COCOON 2001, Lecture Notes Comp. Sc., vol. 2108, 2001, p. 257-262.\\

[4]  P. Caballero-Gil, C. Hernández-Goya, A zero-knowledge identification scheme based on an average-case NP-complete problem, in: Computer Network Security, MMM-ACNS 2003, St. Petersburg, Russia, 
Lecture Notes Comp. Sc., vol. 2776, 2003, p. 289-297.\\

[5] U. Feige, A. Fiat, A. Shamir, Zero knowledge proofs of identity, J. Crypt. 1, 1987, p. 77-94.\\

[6] D. Grigoriev, V. Shpilrain, Authentication schemes from actions on graphs, groups, or rings, Annals of Pure and Applied Logic 162, 2010, p. 194-200.\\

[7] P. Hell, J. Nesetril: On the complexity of H-coloring. J. Comb. Theory, Ser. B 48(1), 1990, p. 92-110 .\\

[8] W. Imrich, S. Klav\v zar, Product Graphs, John Wiley and Sons, 2000.\\
 
[9] L. G. Kachian, A Polynomial algorithm in linear programming, Dokl. Akad. Nauk SSSR 244, 1093-1096, 1979. English translation in Soviet Math. Dokl. 20, 1979, p. 191-194.\\

[10] N. Koblitz, Algebraic Aspects of Cryptography, Vol.3, Algorithms and Computation in Mathematics, Springer-Verlag, 3rd edition, 2004.\\

[11] B. D. McKay, Practical graph isomorphism, Congr. Numer. 30, 1981, p. 45-87.\\

[12] A. Menezes, P. van Oorschot, and S. Vanstone, Handbook of Applied Cryptography, CRC Press, 1996.\\

[13] T. Miyazaki, The complexity of McKay's canonical labeling algorithm, 28, Amer. Math. Soc., 1997, p. 239-256.

[14] Paul T. Darga, Mark H. Lifiton, Karem A. Sakallah, and Igor L. Markov, Exploiting structure in symmetry detection for conference proceedings of the 41th Design automation Conference,  New York, NY, USA, 2004, ACM, p. 530-534.\\

[15] Richard M. Karp, Reducibility among Combinatorial Problems, journal of symbolic logic, vol. 40, No. 4, 1972, p. 85-103.\\

[16] Stephen A. Cook, The complexity of Theorem-Proving Procedures, proceedings third annual ACM symposium on theory of computing, 1971, p. 151-158.\\

[17] G. Tener, Attacks on difficult instances of graph isomorphism: sequential and parallel algorithms, PhD thesis, University of Central Florida, 2009.\\

[18] Tommi Junttila and Petteri Kaski, Engineering an efficient canonical labeling tool for large and sparse graphs, proceedings of the ninth workshop on Algorithm Engineering and Experiments,  SIAM, 2007, p. 135-149.\\

[19] W. Trappe, L. Washington, Introduction to Cryptography with Coding Theory, Prentice Hall, second edition, 2005. \\

\end{document}